\documentclass[12pt]{article}
\usepackage{amsmath,amssymb}

\usepackage{amscd}

\newcommand{\verylongrightarrow}{-\hspace{-0.2cm}-\hspace{-0.2cm}\longrightarrow}

\begin{document} 
\title{Bowen-York Tensors}

\author{Robert Beig \and Werner Krammer\\
Institut f\"ur Theoretische 
Physik der Universit\"at
Wien\\ Boltzmanngasse 5, A-1090 Vienna, Austria\\[1cm]}

\maketitle

\begin{center}
We are honoured and pleased to dedicate this paper to Professor Vince Moncrief
on the occasion of his sixtieth birthday.
\end{center}

\begin{abstract}
There is derived, for a conformally flat three-space, a family of linear 
second-order partial differential operators which 
send vectors into tracefree, symmetric two-tensors. These maps, which are
para\-met\-rized by conformal Killing vectors on the three-space, are such that the 
divergence of the resulting tensor field depends only on the divergence of the
original vector field. In particular these maps send source-free electric
fields into TT-tensors. Moreover, if the original vector field is
the Coulomb field on $\mathbb{R}^3\backslash \lbrace0\rbrace$, the resulting 
tensor fields on $\mathbb{R}^3\backslash \lbrace0\rbrace$ are nothing
but the family of TT-tensors originally written down by Bowen and York.
\end{abstract}

Keywords: TT-tensors, constraints of general relativity

\newcommand{\beq}{\begin{equation}}
\newcommand{\eeq}{\end{equation}}
\renewcommand{\theequation}{\arabic{section}.\arabic{equation}}
\renewcommand{\theequation}{\thesection.\arabic{equation}}

\section{{\bf Introduction}}
\setcounter{equation}{1}

Bowen-York (BY-)tensors  \cite{jbjy80} are a particular class of TT-tensors. 
They have been a useful tool in 
constructing black-hole initial data satisfying the vacuum constraints 
of General Relativity. In the previous work \cite{rb00} one of us (R.B.)
described a scheme which gives BY-tensors a natural place in the
context of the conformal geometry of conformally flat three-spaces.

Recall that the context in which BY-tensors appear is that of finding aymptotically
flat solutions $(g'_{ab},k'_{ab})$ to the vacuum constraints in the maximal,
i.e. $tr k'=0$ - case. One starts with a ``background'' pair $(g_{ab},
k_{ab})$ and tries to solve a quasilinear elliptic PDE called the Lichnerowicz
equation for a conformal factor $\Phi$, so that 
$(g'_{ab} = {\Phi}^4 g_{ab}, k'_{ab} = \Phi^{-2} k_{ab})$  satisfies 
the Hamiltonian constraint. 
In addition, provided that
$k_{ab}$ is TT with respect to $g_{ab}$, the tensor $k'_{ab}$
will be TT with respect to th physical metric $g'_{ab}$.
In the process one assumes boundary conditions
in order for the resulting initial data to be asymptotically flat.

To fix ideas
let us for a start take the initial slice to be 
$\mathbb{R}^3 \backslash \lbrace0\rbrace)$ with flat background
metric $g_{ab}$.
Consider the following 10-parameter set of symmetric, trace-free tensors:  
\begin{eqnarray}
\label{P}
{}^1k_{ab}(\vec P) &=& \frac{3}{2r^2}
[P_a n_b + P_b n_a - (g_{ab} - n_a n_b)(P,n)]\\
\label{S}
{}^2k_{ab}(\vec S) &=& \frac{3}{r^3}
[\epsilon_{cda} S^c n^d n_b + \epsilon_{cdb} S^c n^d n_a]\\
\label{C}
{}^3k_{ab}(C) &=& \frac{C}{r^3}
[3n_a n_b - g_{ab}]\\ 
\label{Q}
{}^4k_{ab}(\vec Q) &=& \frac{3}{2r^4}
[- Q_a n_b - Q_b n_a - (g_{ab} - 5 n_a n_b)(Q,n)],
\end{eqnarray}
where $(\vec P,\vec S,C,\vec Q)$ are constants, $r^2=g_{ab}x^a x^b =
(x,x)$ with $x^a$ cartesian coordinates and $n^a = x^a/r$. The tensors
in (\ref{P}) to (\ref{Q}) are all TT, i.e. they satisfy
\beq
\label{TT}
D^a k_{ab} = 0.
\eeq
The intuitive meaning of these tensors is as follows: The tensor 
${}^1k_{ab}$ in Eq.(\ref{P})
corresponds to a ``source'' at $r=0$ with linear momentum $\vec P$ and zero
angular momentum, these quantities being defined (``measured'') at
$r= \infty$. The tensor ${}^2k_{ab}$ in Eq.(\ref{S}) corresponds to
a source at $r=0$ with zero linear momentum and angular momentum $\vec S$.
The tensor ${}^4k_{ab}$ in Eq.(\ref{Q}) is the result of acting on
${}^1k_{ab}$ by spherical inversion at the sphere $r=1$. To appreciate the
meaning of ${}^4k_{ab}$ one should first understand the role of the
``puncture'' at $r = 0$. Namely, if one solves the Lichnerowicz equation
subject to the boundary condition that $\Phi$ go to one at infinity and
blow up near the origin like $O(1/r)$, the resulting physical initial-data
set is asymptotically flat near infinity AND $r = 0$. The interpretation,
now, of ${}^4k_{ab}$ is that it corresponds to a linear momentum $\vec Q$
measured near the $r = 0$ - infinity. In fact it would conceptually have
been much simpler to treat the two infinities on equal footing and start
on the ``conformally compactified'' manifold $\mathbb{S}^3$ with 
punctures at two antipodal points corresponding to $r = 0$ (``south pole'')
and $r = \infty$ (``north pole'') previously. With any TT-tensor on 
$\mathbb{S}^3$ 
and any puncture
one can associate a 10-tuple of numbers giving the values of certain ``quantities''
of the TT-tensor at these punctures. The physical meaning of these 
quantities is that of linear and angular momentum for the first six. 
The meaning of the remaining four - let us call them the ``$C$-quantity''
and the ``$\vec Q$-quantity`` -  is less clear since these quantities
are not preserved under 
time-evolution. The role of $C$ can be seen e.g. in maximal slicings of the
maximally extended Schwarzschild spacetime \cite{rb98}.
Anyway, using the above prescription in an appropriate sense, 
the tensor ${}^1k_{ab}$ has linear momentum 
$\vec P$ at infinity and $\vec Q$-quantity equal to $\vec P$ at $r = 0$. 
The tensor ${}^2k_{ab}$ has angular momentum $\vec S$ at infinity
and angular momentum equal to $-\vec S$ at $r = 0$. The tensor 
${}^3k_{ab}$ has $C$-quantity equal
to $C$ both at infinity and at $r = 0$. Finally, the tensor ${}^4k_{ab}$ 
has $Q$-quantity equal to $\vec Q$ at infinity and linear momentum 
$\vec Q$ at $r = 0$. All other quantities are zero. 
The above interpretation, 
which has already been given in \cite{rb00}, will be 
reviewed at the beginning of the next section which also contains our
main new result. This consists of a formula, valid on arbitrary
conformally flat three-manifolds,  giving explicit BY-tensors
in our generalized sense in terms of solutions of the Gauss-law constraint 
of electrodynamics. In the final section we mention some applications. 

\section{{\bf The main theorem}}

Suppose $(M,g_{ab})$ is a compact, locally conformally flat three-manifold.
Conformal flatness is equivalent to the condition

\beq
\label{cotton}
D_a L_{bc} = D_b L_{ac},
\eeq  

where 
\beq
\label{Schouten}
L_{ab} = R_{ab} - \frac{1}{4} g_{ab} R,
\eeq
$R_{ab}$ and $R$ being the Ricci and scalar curvature respectively.
A basic object for us is the quantity $X(\xi,\lambda)$ defined for
arbitrary vector fields $(\xi,\lambda)$ by

\beq
\label{X}
X(\xi,\lambda) = 4L_{ab}\xi^a \lambda^b + (D_{[a} \xi_{b]}) (D^{[a} \lambda^{b]})  
- \frac{2}{9}(D\xi)(D\lambda) + \frac{2}{3}[\xi^a D_a D\lambda + 
\lambda^a D_a D\xi ],
\eeq

where the notation $D\xi$ denotes divergence of the vector field $\xi$. Eq.(\ref{X}) 
defines a symmetric, bilinear functional on the space of all vector fields on M. It has 
the further property that it is invariant under conformal rescalings of $M$
in the sense that $\bar{X} = X$ when $\bar{g}_{ab} = \omega^2 g_{ab},\, 
\bar{\xi}^a = \xi, \, \bar{\lambda}^a = \lambda^a$.   

Furthermore, by virtue of Eq.(\ref{cotton}), if
$(\xi,\lambda)$ are both elements of the finite (at most 10-) dimensional 
vector space of conformal Killing vectors (CKV's), $X$ is in fact
constant on $M$. This constant, for a pair $(\xi,\lambda)$, is 
nothing but 1/3 of the Killing metric evaluated on that pair
$(\xi,\lambda)$, viewed as elements of the Lie algebra of CKV´s on $M$. 
We will omit the verification of this fact. The constancy of $X$ on CKV's
will also follow from a more general statement which we will prove shortly.

We assume from now on that $\xi$ is an arbitrary but fixed CKV. Then $X$ can be
viewed as linear functional taking covectors on $M$ into $\mathcal C^{\infty}$-functions 
on $M$. We write $X(\xi,\cdot): \Lambda^{1,2} \to \mathcal C^{\infty,\,0}$. The second
superscript in these spaces refers to conformal weight, when the metric is
given conformal weight 2. Taking, now, the natural $L^2$-adjoint we
obtain a map $j(\xi;\cdot): \mathcal C^{\infty,-3} \to \Lambda^{1,-3}$, since the volume
element on $M$ scales like $\omega^{-3}$. In other words there holds

\beq
\label{rho}
\int_M X(\xi,\lambda) \, \rho \, dV = \int_M g^{ab} \lambda_a \, j_b(\xi; \rho)\, dV.
\eeq

Explicitly we have

\beq
\label{j}
j_b(\xi;\rho) = -D^a(D_{[a} \xi_{b]} \rho)) + \frac{2}{3} (D_b(D\xi))\rho +
\frac{2}{3} D_b D_a (\xi^a \rho) + \frac{2}{9} D_b ((D\xi)\rho) + 
4 L_{ba}\xi^a \rho
\eeq
Next consider the equation

\beq
\label{BY}
D^ak_{ab} = j_b(\xi;\rho)
\eeq

on $\mathbb{R}^3$ with $\rho = 4 \pi \delta_{0}$, the delta distribution 
concentrated at the origin.
Then the ten TT-tensors given in Eq.'s (\ref{P})- (\ref{Q}) are 
solutions of Eq.(\ref{BY}) when the $\xi$ 's run through CKV's
on $\mathbb{R}^3$. Alternatively, after conformal compactification,
we can view these tensors as solutions of Eq.(\ref{BY}) on $\mathbb{S}^3$, with
$\rho$ given by $\delta_{north pole} - \delta_{south pole}$. This choice of 
punctures is dictated by the following observation: If one contracts
Eq.(\ref{BY}) with another CKV, say $\eta$, and integrates over $M$, one
finds that

\beq
\label{zero}
0 = X(\xi,\eta) \int_M \rho \, dV.  
\eeq   
Since the full conformal group acts on $\mathbb{S}^3$ by conformal isometries 
and the Killing metric of the conformal group is non-degenerate, the
integral of $\rho$ over $M$ has to vanish. Furthermore the right-hand side of
(\ref{zero}), when integrated only over a region excluding one of the
punctures, gives rise to the set of constants $(\vec P,\vec S,C, \vec Q)$, mentioned in 
the Introduction.
These things are more fully explained in \cite{rb00}.

We now come to the insight on which the main result of this paper
is based: since $X(\xi,\lambda)$
is constant when $\lambda$ is a CKV, the gradient of $X$, for general
$\lambda$, can only depend on $l_{ab} = (CK)_{ab}(\lambda)$, where $CK$ 
is the conformal Killing operator acting on $\lambda$,
i.e. the r.h. side of Eq.(\ref{lambda}).
Consequently there exists a linear operator 
$i: \Sigma^{2,2} \to \Lambda^{1,0}$ - where 
$\Sigma^{2,2}$ denotes covariant,
valence-two, symmetric, trace-free tensors on $M$ scaling like $\omega^2$ -
so that

\beq
\label{gradient}
D_a X(\xi,\lambda) = i_a(\xi;CK(\lambda)) 
\eeq
In view of the definition of $X$ and the formulae of the Appendix, the map  $i$
has to be a second-order partial differential operator in its second
argument. Taking the $L^2$-product of Eq.(\ref{gradient}) with some 1-form $E$ in
$\Lambda^{1,-1}$, using Eq.(\ref{rho}) on the left-hand side and integrating
the right-hand side by parts twice, we see that there
exists a second-order partial differential operator 
$BY(\xi;\cdot): \Lambda^{1,-1} \to \Sigma^{2,-1}$
so that

\beq
\label{divE}
\int_M (divE) \, X(\xi,\lambda) \,dV = \int_M g^{ab} g^{cd} \,
(BY)_{ac}(\xi;E) \, (CK)_{bd}(\lambda) \, dV. 
\eeq
Note that the operator $div: \Lambda^{1,-1} \to C^{\infty,-3}$
is minus the $L^2$-adjoint of $grad$. Using Eq.(\ref{rho}) on the left-hand side of 
Eq.(\ref{divE}) and partially integrating the right-hand side of Eq.(\ref{divE}) 
once more, there finally follows that

\beq
\label{final}
\int_M g^{ab} \lambda_a \, j_b(\xi;divE) \, dV =
\int_M g^{ab} (Div\circ BY)_a(\xi;E) \, \lambda_b \, dV 
\eeq

Here $Div$ is the divergence-operator
mapping $\Sigma^{2,1}$ into $\Lambda^{1,-3}$. Since Eq.(\ref{final}) holds
for arbitrary fields $\lambda$, we have obtained the result that the operator
$BY$ satifies the identity

\beq
\label{thm}
Div \circ BY(\xi;E) = j(\xi;divE). 
\eeq
(Eq.(\ref{thm}) is of course an entirely local statement: to derive it 
the integrations by parts have been used for convenience.) In other words
we have obtained the following result

{\bf{Theorem:}} Suppose that $E$ satisfies $divE = \rho$. Then the symmetric,
trace-free tensor $k_{ab}$ defined by 

\beq
\label{1}
k_{ab} = (BY)_{ab}(\xi;E)
\eeq

satisifies the Bowen-York equation (\ref{BY}), namely

\beq
\label{2}  
D^a k_{ab} = j_b(\xi;\rho).
\eeq

In order for this result to be useful it remains to give the explicit expression
for $BY$. We first have to find $i_a$ by taking the gradient of $X$. Using
Eq.'s (\ref{lambda}), (\ref{curl}) and (\ref{DDdiv}) together with the remark 
following it, we find after a lengthy but straightforward calculation

\begin{eqnarray}
\label{i}
\lefteqn{ i_a(\xi; l) = \xi^b[(-2)\Delta l_{ab} + 4 D_{(a}D^c l_{b)c} - 
g_{ab} D^c D^d l_{cd} - {} }
\nonumber\\
 && {} \quad\quad - 3 g_{ab} \, l^{cd} L_{cd} + 6 L_a \,^c l_{bc} + 
10 L_ b \,^c l_{ac}] + {} 
\nonumber\\ 
 && {} \quad\quad + 2 (D^{[b}\xi^{c]})(D_{[b} l_{c]a}) + 
\frac{2}{3} (D^b D\xi) l_{ab}
\end{eqnarray}

Recall that $BY$ is the $L^2$-adjoint of $i$. Using that CK($\xi$) is zero
we obtain by another lengthy computation that

\begin{eqnarray}
\label{lengthy}
\lefteqn{ (BY)_{ab}(\xi;E) = - \xi^c D_{(a}D_{b)}E_c - 2 \xi_{(a}\Delta E_{b)} +
g_{ab} \xi^c \Delta E_c + 2 \xi_c D_{(a} D^c E_{b)} + {}}
\nonumber\\ 
 && {} + 2 \xi_{(a}D_{b)}(DE) - \frac{4}{3} g_{ab} \xi^c D_c(DE) + 
 \frac{4}{3} (D\xi)D_{(a} E_{b)} - 8 F_{c(a} D^c E_{b)} + 4 F^c \,_{(a} D_{b)} E_c - {}
\nonumber\\
 && {} - \frac{4}{9} g_{ab} (D\xi)(DE) + 4 g_{ab} F^{cd} D_c E_d + 4 E_{(a} D_{b)} D\xi - 
\frac{4}{3} g_{ab} E^c D_c (D\xi) + 9 \xi^c L_{c(a} E_{b)} + {}
\nonumber\\
 && {} + \xi_{(a} L_{b)c} E^c + 4 L \, \xi_{(a} E_{b)} - 2 g_{ab} \,L \, \xi^c E_c,  
\end{eqnarray}
where $L = g^{ab}L_{ab}$ and $F_{ab} = D_{[a} E_{b]}$. 

The space of TT-tensors is infinite-dimensional. This
is due to the fact that the operator Div is underdetermined-elliptic. Thus the
equation (\ref{j}) has many solutions.  If M is asymptotically flat and
appropriate boundary conditions are assumed or when M is compact, a unique 
solution can be found by means of the York-decomposition. This consists
of the observation that $\Sigma^2$ splits into an $L^2$-orthogonal 
sum of tensors which
are TT and ``longitudinal ones'', i.e. ones of the form $k = CK(W)$ for
some covector $W$. Note
that this splitting breaks conformal invariance (hence the second superscript 
in $\Sigma$ was omitted). Since
$Div \circ CK$ is elliptic with null space solely the CKV's, Eq.(\ref{BY})
has a unique longitudinal solution. Similarly the equation $div E = \rho$
has a unique longitudinal solution given by $E = grad \,G$ for some scalar
function $G$. One is thus led to the question whether $BY$ maps longitudinal
solutions of the respective spaces into one another. There is an affirmative
answer to this question in the case where $g_{ab}$ is a space form, i.e. has
constant curvature rather
than merely being conformally flat. Namely, in that case,  there exists an operator 
$h(\xi;\cdot): C^{\infty} \to \Lambda^1$ such that
$CK \circ h(\xi,G) = BY(\xi, grad \,G)$. The operator $h$ appears already 
in the work \cite{rb00}. The restriction to constant curvature for the compactified 
manifold leaves one with the case of $\mathbb{T}^3$, $\mathbb{S}^3$ and quotients thereof. 
The present work was motivated by the desire to do similar things on 
$\mathbb{S}^2 \times \mathbb{S}^1$, and this finally led us to the operator
$BY$.  

We can now clarify 
the connection between our map $BY$ and the Bowen-
York expressions in (\ref{P})-(\ref{Q}). Let us take M to be $\mathbb{R}^3$.
The CKV's are explicitly given by

\begin{eqnarray}
\label{trans}
{}^1 \xi^a (\vec{\pi}) &=& \pi^a\\
\label{rot}
{}^2 \xi^a (\vec{\sigma}) &=& \epsilon^a \, _{bc}\sigma^b x^c\\
\label{dil}
{}^3 \xi^a (\zeta) &=& \zeta x\\ 
\label{boost}
{}^4 \xi^a (\vec{\gamma}) &=& (x,x) \gamma^a - 2(x,\gamma) x^a, 
\end{eqnarray}
where $\pi^a,\sigma^a,\zeta,\gamma^a$ are constants.
Let $E$ be the Coulomb solution of $div E = 4 \pi \delta_0$, i.e.

\beq
\label{coulomb}
E = \frac{n}{r^2}.
\eeq

Then

\begin{eqnarray}
\label{P1}
{}^1k (\vec P) &=& -\frac{1}{2} BY({}^4 \xi (\vec P),E)\\
\label{S1}
{}^2k (\vec S) &=& - BY({}^2 \xi (\vec S),E)\\
\label{C1}
{}^3k (C) &=& - BY({}^3\xi(C),E)\\
\label{Q1}
{}^4k (\vec Q) &=& \frac{1}{2} BY({}^1 \xi (\vec Q),E)
\end{eqnarray}

The main result of this paper is part of a more elaborate scheme. To describe
it recall that the operators $grad$ and $div$ and are respectively the left 
and right end of the de Rham complex on M, which has the operator $rot$, given
by $(rot \,\mu)_a = \epsilon _a\;^{bc} D_b \mu_c$, in the middle. There exists
a similar elliptic complex, introduced in \cite{rb97}, with $CK$ and $Div$
at the left and right end, respectively. The operator in the middle is 
(1/2 of) a
third-order partial differential operator called $H$ in \cite{rb97} which
is the linearization, at the conformally flat metric afforded by $g_{ab}$, 
of the Cotton-York tensor applied to trace-free symmetric tensors. These two elliptic
complexes are related by the following commutative diagram:

\begin{equation}
\begin{array}{ccccccccccc}
0 & \rightarrow & {\mathcal C}^{\infty,0} & \stackrel{grad}{\verylongrightarrow} & \Lambda^{1,0} & \stackrel{rot}{\verylongrightarrow} & 
     \Lambda^{1,-1} & \stackrel{-div}{\verylongrightarrow} &  {\mathcal C}^{\infty,-3} & \rightarrow & 0     \\
  &  & \uparrow    &  & \uparrow & & \uparrow & & \uparrow & & \\[-1ex]
  &  \mbox{{\scriptsize $X(\xi,\cdot)$}} \hspace{-4ex}& | & \mbox{{\scriptsize $i(\xi;\cdot)$}}\hspace{-6ex}   
& | & \mbox{{\scriptsize $\xi \lrcorner\;\cdot$}} \hspace{-8ex} & | &\mbox{{\scriptsize $\xi\lrcorner\;\cdot$}}\hspace{-10ex}& | & & \\[-1ex]
  &  & | & & | & & | & & | & & \\    
0 & \rightarrow &  \Lambda^{1,2} & \stackrel{CK}{\verylongrightarrow} & \Sigma^{2,2} & \stackrel{\frac{1}{2}H}{\verylongrightarrow} & 
     \Sigma^{2,-1} & \stackrel{-Div}{\verylongrightarrow} &  \Lambda^{1,-3} & \rightarrow & 0   \\  
  &  & \uparrow    &  & \uparrow & & \uparrow & & \uparrow & & \\[-1ex]
  &  \mbox{{\scriptsize $\xi \otimes \;\cdot$}} \hspace{-4ex}& | & \mbox{{\scriptsize $\tau(\xi\otimes\cdot)$}}\hspace{-6ex}   
& | & \mbox{{\scriptsize $BY(\xi;\cdot)$}} \hspace{-8ex} & | & \mbox{{\scriptsize $j(\xi,\cdot)$}} \hspace{-10ex}& | & & \\[-1ex]
  &  & | & & | & & | & & | & & \\    
0 & \rightarrow & {\mathcal C}^{\infty,0} & \stackrel{grad}{\verylongrightarrow} & \Lambda^{1,0} & \stackrel{rot}{\verylongrightarrow} & 
     \Lambda^{1,-1} & \stackrel{-div}{\verylongrightarrow} &  {\mathcal C}^{\infty,-3} & \rightarrow & 0  
\end{array}
\end{equation}

The two lower-left and upper-right vertical maps are algebraic: The one on the far left is simply 
multiplication of $\mathbb{C}^{\infty}$-functions by $\xi$. The next one is the tensor
product with $\xi$ combined with taking the symmetric, trace-free part. The right upper arrows denote
contraction with $\xi$. Note that there is, in this diagram, a symmetry: If one simultaneously permutes 
top and bottom and left and right, one obtains the adjoint map of the original one.

After completion of this work we learnt that the above structure should be a 
special case of a general construction known as the ``Bernstein-Gelfand-Gelfand'' resolution
\cite{ac99}. For a more pedagogical exposition of this material we refer to the works
\cite{me99} and \cite{me00}.

\section{Applications}

We are interested in finding BY-tensors on manifolds with nontrivial topology. Take
for example the background conformal structure  to be the standard one on 
$\mathbb{S}^2 \times \mathbb{S}^1(a)$, 
i.e. the unit-two sphere times the circle with radius $a$. Solving
the Lichnerowicz equation with one puncture and vanishing extrinsic curvature,
one obtains the Misner womhole (see \cite{cm60}), with the radius $a$ determining  
the ``distance'' between the wormholes in units of the total mass. Taking, on that
manifold, two punctures which lie in the same fibre of the $\mathbb{S}^2$-factor
and antipodally on the circle, one obtains Einstein-Rosen data with two bridges.
(For a nice review of these initial data see \cite{dg98}.)
The aim is to construct non-time symmetric generalizations of such  initial data.
(For known such data and their use in Numerical Relativity see the review
in \cite{gc00}.) We do this 
by taking for the background extrinsic curvature BY-data as described  
in the present work. One is then left with the task of solving the Gauss law
$div E = \rho$ with appropriate delta function sources 
and applying to the result the BY-map found in this paper. 
In this way one is able to give new expressions for existing BY-type data 
and also to construct new ones. Details will be given in the forthcomiing PhD-
thesis by one of us (W.K.)

{\bf{Acknowledgments}}: One of us (R.B.) thanks Mike Eastwood for pointing out to him
the work of A.Cap and collaborators and Andreas Cap for explanations.

\begin{appendix}
\section{Appendix}
\setcounter{equation}{0}

In this appendix we collect a few formulae which are important in the text.
Let $(M, g_{ab})$ be a three-manifold. Define the Schouten 
tensor by

\beq
\label{sch}
L_{ab} = R_{ab} - \frac{1}{4} g_{ab} R.
\eeq

In three dimensions the Riemann tensor can be expressed in terms of the Ricci
tensor by means of

\beq
\label{riem}
R_{abcd} = 2L_{c[a} g_{b]d} - 2L_{d[a}g_{b]d}.
\eeq

Suppose $\xi$ is a conformal Killing vector. There is then a well-known 
way by means of which one can write down a sequence of integrability
conditions, sometimes called ``conformal Killing transport equations''
which imply that $\xi$ is given uniquely by its ``conformal Killing data''
at some arbitrary point in $M$. These Killing data are furnished by 

\beq
\label{data}
(\xi_a, D_{[a} \xi_{b]}, D\xi, D_a D\xi),
\eeq

where $D\xi$ and $D_aD\xi$ are respectively the divergence and the 
gradient of the divergence of $\xi$. What we do here is to perform
the same procedure on the equation

\beq
\label{s}
(CK)_{ab} (\lambda) = D_{(a} \lambda_{b)} - \frac{1}{3} g_{ab} D\lambda = l_{ab},
\eeq
which is simply the definition of the conformal Killing form for an
arbitrary vector field $\lambda$. Equivalently we can write

\beq
\label{lambda}
D_a\lambda_b = l_{ab} + D_{[a}\lambda_{b]} + 
\frac{1}{3} g_{ab} D\lambda
\eeq

Using the Ricci identity and (\ref{riem}) we obtain 

\beq
\label{curl}
D_a D_{[b}\lambda_{c]} = - 2 L_{a[b}D_{c]}(D\lambda) - 2g_{a[b}L_{c]d}
\lambda^d - \frac{2}{3}g_{a[b}D_{c]}D\lambda + 2 D_{[b}l_{c]a}
\eeq

and

\begin{eqnarray}
\label{DDdiv}
\lefteqn{ D_aD_bD\lambda = -3 \mathcal{L}_{\lambda}L_{ab} + 
18L^c\,_{(a} l_{b)c} - \frac{9}{2}g_{ab}l^{cd}L_{cd} -{} } 
\nonumber\\
& & {}\quad\quad\quad - 3 \Delta l_{ab} + 6D_{(a}D^c l_{b)c} - \frac
{3}{2}g_{ab}D^cD^d l_{cd}
\end{eqnarray}

Note that the first term on the r.h.side of Eq.(\ref{DDdiv}) can be expressed
in terms of $l_{ab}$ and the would-be conformal Killing data of $\lambda$ if 
$\lambda$ is a CKV. 

\end{appendix}

\end{document}